\documentclass[a4paper,twoside]{article}
\usepackage{caption}
\usepackage{subcaption}
\usepackage{subcaption}
\usepackage{calc}
\usepackage{amssymb}
\usepackage{amstext}
\usepackage{amsmath}
\usepackage{array}    
\usepackage{float}    
\usepackage{url}      
\usepackage{epsfig}  
\usepackage{xurl}
\usepackage{paralist}
\usepackage{tabularx} 
\usepackage{booktabs}

\usepackage{placeins} 
\usepackage{xcolor}   
\usepackage{todonotes}[color=red;backgroundcolor=red] 
\usepackage{amsthm}
\usepackage{multicol}
\usepackage{pslatex}
\usepackage{apalike}
\usepackage{algorithm}

\usepackage{algpseudocode}
\algrenewcommand\algorithmicrequire{\textbf{Input:}}
\algrenewcommand\algorithmicensure{\textbf{Output:}}

\usepackage[bottom]{footmisc}
\usepackage{hyperref}
\usepackage{balance}
\usepackage{SCITEPRESS}
\usepackage{flushend}

\begin{document}


\title{Toward Accessible Mobile Money: A Voice-Driven, Biometrically Secured USSD Automation Framework for Visually Impaired Users}

\author{
      \authorname{Sunday Ajayi \sup{1}\orcidAuthor{0009-0002-2076-916X},
            Babatunde Eric Olatunji\sup{1}\orcidAuthor{0009-0002-5304-3146}
            and Eric Umuhoza\sup{1}\orcidAuthor{0000-0002-2451-8897}}
      \affiliation{\sup{1}Carnegie Mellon University Africa, Kigali, Rwanda}
      \email{\{sajayi, bolatunj, eumuhoza\}@andrew.cmu.edu}
}


\abstract{
Financial inclusion has expanded significantly across Africa through mobile money services delivered primarily via USSD technology. However, visually impaired individuals continue to face accessibility and security barriers when conducting financial transactions. Current USSD systems are not designed for non-visual interaction, forcing users to rely on third-party assistance even for PIN entry, thereby increasing fraud exposure and reducing transaction confidence.
Although alternative assistive technologies  such as screen readers exist, they are not  compatible with USSD operations, often causing sessions to time out before the user can complete a transaction.
This paper presents an Android-based intelligent middleware that automates USSD transactions, integrates biometric-secured PIN injection, and introduces a privacy-preserving screen-dimming mechanism: \emph{Blackout Mode}. The system leverages Android Accessibility Services, hardware-backed Keystore security, and on-device natural language parsing to enable independent, secure voice-based mobile money access.
We show that the proposed solution improves task success rates from 65–75\% to more than 90\% and reduces transaction completion time from 40–60 seconds to 12--15 seconds, while also improving perceived security.
}

\onecolumn \maketitle \normalsize \setcounter{footnote}{0} \vfill

\section[Introduction]{\uppercase{Introduction}}

Financial inclusion refers to access to affordable and responsible financial services, including payments, savings, credit, and insurance, as defined by the World Bank \cite{WorldBank2022}. Despite the rapid growth of digital financial services in Africa, people with disabilities (PWDs) continue to face significant exclusion. Across Africa, mobile money has significantly expanded inclusion through  Unstructured Supplementary Service Data (USSD) based services from operators such as MTN, Safaricom (Mpesa), and Airtel.

In Rwanda, reports from the United Nations Development Program (UNDP Rwanda) show that PWDs, particularly visually impaired individuals, remain disproportionately excluded from financial services despite national inclusion policies \cite{UNDPRwanda2021}. Visual impairment, one of the most common disabilities, presents unique challenges in accessing digital financial platforms. Addressing these challenges is essential for achieving universal financial inclusion \cite{eprn2021}.

USSD, highlighted by the GSMA as a low-tech high-impact financial channel \cite{GSMA2020}, remains the dominant transaction medium in East Africa due to its compatibility with low-end devices and zero dependence on internet connectivity. In 2019, the mobile money transaction made up to about 64.15\% of all transactions made worldwide that was from Sub-Saharan Africa and the Middle East and Northern Africa \cite{statista_mobile_money_africa}

It works on both basic phones and smart phones without requiring internet connectivity, which makes it especially important for low-income and rural populations. However, USSD systems were not designed for non-visual interaction, which makes them unable to navigate using assistive technologies like screen readers.

As a result, many visually impaired users rely on third-party assistance to complete mobile money transactions, exposing sensitive information such as PINs and significantly increasing their risk of fraud \cite{Alajarmeh2021}.

Although widely adopted and promoted by the GSMA as a low-cost, high-impact digital channel, USSD technology was not designed with accessibility in mind. Consequently, visually impaired users face several challenges:
\begin{inparaenum}[(i)]
\item strict session timeouts (typically 60--120 seconds) \cite{GSMA2020};
\item limited compatibility with screen readers \cite{billah2017};
\item exposure of PINs during assisted transactions; and
\item increased vulnerability to fraud \cite{Muuo2024}.
\end{inparaenum}

These limitations create a critical trade-off between security and accessibility, where improved usability often comes at the expense of privacy and user independence.

In this paper, we propose an Android-based intelligent middleware system that enables visually impaired users to conduct mobile money transactions independently and securely, without requiring external assistance. The system automates USSD interactions, integrates biometric-secured PIN authentication, and incorporates a privacy-preserving screen-dimming mechanism.

Our study is guided by the following research questions:

\begin{itemize}
    \item \textbf{RQ1:} How do privacy-enhancing features, such as screen dimming (blackout mode), influence perceived security, trust, and usability among visually impaired users?

    \item \textbf{RQ2:} How does automated USSD interaction compare with traditional manual USSD usage in terms of cognitive load, error rates, and user independence?

    \item \textbf{RQ3:} What are the key barriers and enablers to adopting secure USSD automation solutions among visually impaired mobile money users in low-resource settings?

\end{itemize}

\section{\uppercase{Background}}
This section analyzes the accessibility challenges existing in USSD systems, particularly session time limitations, and reviews existing interventions designed to mitigate these barriers, including digital accessibility solutions and regulatory frameworks implemented by government authorities.

\subsection{Challenges in USSD Accessibility}

Existing studies highlight several key challenges faced by visually impaired users when interacting with USSD systems: 
\begin{itemize}
    \item \emph{Session Timeout Constraints.} USSD sessions typically enforce strict time limits ranging from 60 to 120 seconds \cite{GSMA2020}. Visually impaired users often require more time to navigate menus using assistive technologies, leading to frequent session expirations and failed transactions \cite{billah2017};
   \item \emph{Screen Reader Limitations.} Although screen readers such as TalkBack and VoiceOver support smartphone accessibility, they are largely incompatible with USSD interfaces \cite{Alajarmeh2021}. In general, the USSD menus are linear, text-based, and lack structural semantics, making them difficult for screen readers to interpret accurately \cite{billah2017}. This limitation is not unique to USSD interfaces. In fact, Oh et al. highlighted that many systems are not designed with screen reader workflows in mind \cite{oh2021imageaccessibility}; and 
   \item \emph{Security and Privacy Risks.} USSD transactions often involve privacy sensitive steps such as PIN entry or OTP verification. Visually impaired users frequently rely on third-party assistance to complete these steps, which can compromise privacy and increase exposure to fraud \cite{myjoyonline2022}, \cite{Muuo2024}.
   
\end{itemize}

\subsection{Accessibility Interventions}

\subsubsection{Policy and Regulatory Efforts}
The International Telecommunication Union (ITU), the United Nations specialized agency for information and communication technologies, underscores the need to integrate accessibility considerations into ICT laws and regulatory frameworks \cite{itu_ict_accessibility_policy_2024}.

Several countries, including Ghana, Uganda, Rwanda, Palestine and Sierra Leone, have introduced policies aimed at improving accessibility to financial systems for PwDs. These efforts include 
   \emph{adoption of Braille-enabled ATMs} and \emph{audio-guided banking systems}.
However, these policies primarily address physical banking infrastructure and do not adequately cover digital financial channels.

\subsubsection{Mobile Application-based Solutions}
Some telecommunications providers have begun deploying digital innovations aimed at enhancing accessibility. For instance, Safaricom introduced an interactive voice response (IVR) interface within M-Pesa, enabling visually impaired users to conduct transactions through audio-guided prompts \cite{safaricom2017}.

Similarly, financial institutions are increasingly integrating accessibility features into mobile banking platforms to enhance financial access and usability for visually impaired users \cite{wanjiku2025mobilebanking}.

Although these solutions represent meaningful progress, they remain constrained by important limitations: they typically require access to smartphones and depend on reliable internet connectivity. Consequently, such approaches risk excluding users who rely on basic feature phones or who lack consistent internet access, precisely the populations that depend mainly on USSD-based services.

\section{\uppercase{Proposed System}}
The proposed system is an Android-based middleware solution designed to enhance accessibility and security in USSD-based financial transactions for visually impaired users. The system operates as an intermediary layer between the user and the existing USSD infrastructure of the telecom.

\begin{figure*}[h]
\centering
\includegraphics[width=0.7\linewidth]{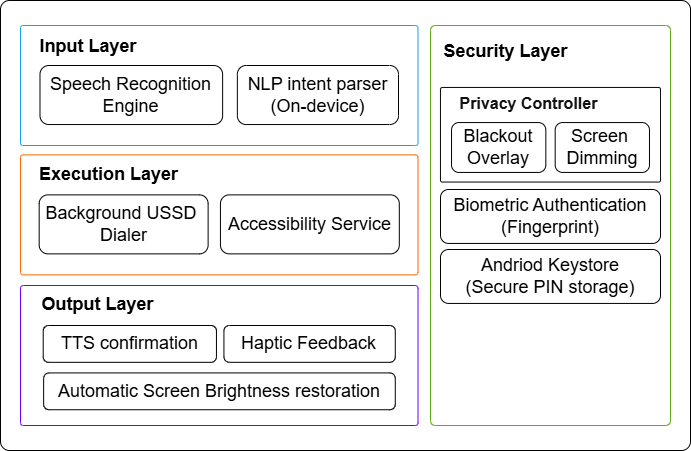}

\caption{Layered system architecture showing the data flow and interaction between input processing, USSD execution, security enforcement, and accessible output modules}
\label{fig:system_architecture}
\end{figure*}

To effectively manage the complexity of automating USSD sessions while maintaining rigorous privacy standards, the system is designed using a modular, layered architecture depicted in Figure \ref{fig:system_architecture}.
The interaction begins with the capture and parsing of natural language \verb|Input Layer|, which triggers the automated interaction of the mobile network \verb|Execution Layer|. Before sensitive operations are completed, the system enforces a strict privacy gateway and credential retrieval \verb|Security Layer|, culminating in multimodal user confirmation \verb|Output Layer|. This separation of concerns ensures that Android APIs, on-device machine learning models, and hardware-based security components work together smoothly without exposing sensitive data between modules.

\subsection{Input Layer}

The Input Layer helps in capturing user interaction and transforming it into structured transaction intent. It serves as the entry point of the system and comprises the following two sub-systems:

\begin{enumerate}
    \item \emph{Speech Recognition Engine:} It captures spoken input from the user and converts voice commands (e.g., ``Send 5000 to 0722334444'') into textual data.

\item \emph{On-device NLP Intent Parser:} It processes the transcribed text using the on-device NLP processing to extract structured transaction details such as action type (e.g., send money), amount, and recipient information. This processing is performed on-device to ensure privacy, reduce latency, and maintain offline functionality.
\end{enumerate}

\subsection{Execution Layer}

The Execution Layer is responsible for interacting with the telecom infrastructure and automating USSD-based financial transactions such as fund transfers, airtime top-up, and balance checks. The layer comprises the following two sub-systems:
\begin{enumerate}
  
\item \emph{Background USSD Dialer:} This automatically initiates USSD sessions based on the parsed intent without requiring manual user dialing. It constructs the appropriate USSD string and executes it in the background.

\item \emph{Accessibility Service --- UI Interaction and Response Extraction:} This monitors USSD pop-up interfaces generated by the mobile network. It extracts textual content from these interfaces and passes it to the application for processing. This enables the system to interpret and respond to USSD prompts in real time without visual interaction.
\end{enumerate}

\subsection{Security Layer}
The Security Layer ensures confidentiality, integrity, and user authentication throughout the transaction process. It is the core trust component of the system. It comprises the following sub-systems:

\begin{enumerate}
\item \emph{Privacy Controller --- Screen Dimming and Blackout Overlay:} This activates a blackout mode during sensitive operations such as PIN entry by reducing screen brightness to zero and overlaying a full-screen black interface. This prevents visual exposure of sensitive information.

\item \emph{Biometric Authentication---Fingerprint:} uses the device’s biometric system (via BiometricPrompt API) to verify the identity of the user before allowing access to sensitive operations such as PIN retrieval.

\item \emph{Secure PIN Storage --- Android Keystore:} This stores the user’s encrypted PIN in a hardware-backed Trusted Execution Environment (TEE). This ensures that the PIN is never exposed in plaintext and cannot be accessed by unauthorized applications or processes.
\end{enumerate}

\subsection{Output Layer}

The Output Layer is responsible for providing feedback to the user in accessible and multimodal formats, ensuring usability for visually impaired users. It consists of the following sub-systems:

\begin{enumerate}
   
\item \emph{Text-to-Speech (TTS) Engine:} This converts system messages, USSD responses, and transaction confirmations into spoken audio output.

\item \emph{Haptic Feedback Module:} It provides vibration-based feedback to indicate different system states such as success, failure, or required user action.

\item \emph{Automatic Screen Brightness Restoration:} This restores the device’s screen brightness to its original state after transaction completion, ensuring normal device usage resumes seamlessly.
\end{enumerate}

\section[Implementation]{\uppercase{Implementation}}
This section details the practical realization of the proposed middleware, structured  according to the four-layer system architecture defined previously in Figure~\ref{fig:system_architecture}. To clearly illustrate the system’s operational logic without relying on low-level code snippets, the implementation of each layer (Input, Execution, Security, and Output) is presented through high-level algorithmic workflows. These conceptual workflows demonstrate the step-by-step orchestration of native Android components, including how spoken intents are parsed, how USSD dialogs are programmatically navigated in the background, how the Android Keystore enforces cryptographic security during PIN injection, and how transaction states are finalized using multimodal feedback.


\subsection{Prototype Development}

The prototype was developed natively for the Android platform using Kotlin and the Jetpack Compose declarative user interface toolkit. The system is designed as an intelligent middleware layer that interfaces with the existing USSD infrastructure while leveraging the capabilities of the Android system-level for automation, accessibility, and security.

At the input layer (see Algorithm~\ref{aglo:algo1}), the system incorporates a speech-to-intent processing module that captures user voice commands and translates them into structured financial intents, along with relevant metadata such as recipient and transaction amount.

To enable real-time USSD monitoring and interaction, users are required to grant permission to a custom Accessibility Service through the native Android settings interface. This permission allows the application to programmatically observe, interpret and interact with USSD dialog windows, including automatic text extraction, input injection, and navigation across menu layers.

The execution layer, as implemented in Algorithm~\ref{aglo:algo2}, automates USSD sessions by dynamically reading on-screen content, triggering text-to-speech feedback, and performing context-aware actions such as PIN entry and menu selection. This significantly reduces the need for manual navigation and minimizes user input errors.

To ensure secure transaction handling, the system integrates a privacy-preserving security layer, implemented through Algorithm~\ref{aglo:algo3}, which requires biometric authentication before retrieving and decrypting the user’s stored PIN. During sensitive operations, visual output is suppressed using a blackout overlay and reduced screen brightness to prevent shoulder surfing.

Finally, the output layer, implemented through Algorithm~\ref{aglo:algo4}, provides multimodal feedback by delivering transaction results through synthesized speech and haptic signals, which ensures that users receive clear and immediate confirmation of transaction outcomes.

The onboarding and setup process was designed to be fully navigable using Android’s native screen reader, ensuring complete accessibility compliance from installation. This guaranties that visually impaired users can independently configure and operate the system without external assistance.

\begin{algorithm}[htbp]
\small
\caption{Input Layer: Speech-to-Intent Processing}
\label{aglo:algo1}
\begin{algorithmic}[1]

\Require User voice stream $S$
\Ensure Target intent $I$, metadata $M = \langle target, amount \rangle$
\Statex

\State $Recognizer \gets \textsc{CreateSpeechRecognizer}()$
\State $match \gets Recognizer.\textsc{Listen}(S)$
\State $cmd \gets \textsc{Lowercase}(match)$
\State $num \gets \textsc{ParseSpokenDigits}(cmd)$

\If{$cmd$ contains ``balance''}
    \State \Return $\langle \text{BAL\_CHECK}, 0 \rangle$
\EndIf

\If{$cmd$ contains ``send'' \textbf{or} ``transfer''}
    \State $\langle target, amount \rangle \gets \textsc{ExtractContext}(cmd, num)$
    \If{$target \neq \emptyset$ \textbf{and} $amount \neq 0$}
        \State \Return $\langle \text{TRANSFER}, \{target, amount\} \rangle$
    \EndIf
\EndIf

\end{algorithmic}
\end{algorithm}

\begin{algorithm}[htbp]
\small
\caption{Execution Layer: USSD \& UI Automation}
\label{aglo:algo2}
\begin{algorithmic}[1]

\Require USSD code $C$, user PIN $P$

\State \textsc{Dial}($C$)

\While{\textsc{SessionActive}()}
    \State $Node \gets \textsc{GetRootInActiveWindow}()$
    \State $UI\_Text \gets \textsc{ExtractText}(Node)$
    \State \textsc{TriggerTTS}($UI\_Text$)

    \If{$Node$ is editable \textbf{and} PIN required}
        \State \textsc{InjectText}($Node, P$)
        \State \textsc{PerformClick}(``Send'')
    \EndIf

    \If{\textsc{IsTransactionComplete}($UI\_Text$)}
        \State \textsc{TerminateSession}()
        \State \textbf{break}
    \EndIf
\EndWhile

\end{algorithmic}
\end{algorithm}

\begin{algorithm}[htbp]
\small
\caption{Output Layer: Multimodal Feedback Controller}
\label{aglo:algo3}
\begin{algorithmic}[1]

\Require Result status $S$, transaction text $T$

\State \textsc{SpeakDirectly}($T$)

\If{$S = \text{SUCCESS}$}
    \State \textsc{Vibrate}(\textit{pattern: SHORT})
\Else
    \State \textsc{Vibrate}(\textit{pattern: LONG\_PULSE})
\EndIf

\State \textsc{Broadcast}(``TRANSACTION\_FINISHED'')
\State \textsc{SetScreenBrightness}(\textsc{Normal})
\State \textsc{ClearWindowFlags}(\textsc{FlagFullscreen})

\end{algorithmic}
\end{algorithm}

\begin{algorithm}[htbp]
\caption{Security Layer: Privacy \& Vault Controller}
\label{aglo:algo4}
\begin{algorithmic}[1]

\Require Stored encrypted PIN $E$, session request $R$

\State \textsc{ShowBiometricPrompt}(``Identity Verification'')

\If{\textsc{AuthenticationSucceeded}()}
    \State $Key \gets \textsc{GetKey}($``PIN\_Alias''$)$
    \State $P \gets \textsc{Decrypt}(E, Key)$
    
    \State \textsc{ShowBlackoutOverlay}(\textsc{Opaque})
    \State \textsc{SetScreenBrightness}(0.0)
    
    \State \textsc{ExecuteSession}($R, P$)
\Else
    \State \textsc{AbortSession}(``User not verified'')
\EndIf

\end{algorithmic}
\end{algorithm}

\subsection{Deployment}

\begin{figure*}[h]
\centering
\includegraphics[width=.9\linewidth]{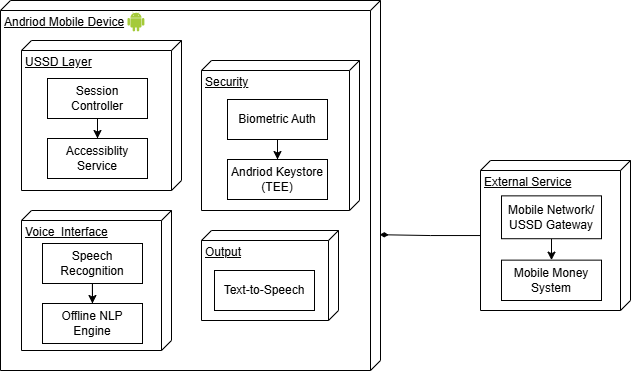}

\caption{Deployment Architecture of a Secure USSD-Based Mobile Financial Services Application on Android. The diagram illustrates the five functional layers within the Android device boundary --- USSD Layer, Voice Interface, Security, and Output --- and their unidirectional communication with the External Service block comprising the Mobile Network/USSD Gateway and Mobile Money System.}
\label{fig:deployment_architecture}
\end{figure*}

The system is deployed within the Android operating system Figure \ref{fig:deployment_architecture}. This choice is driven by the specific technical 
requirements of the solution, which depend on deep system-level integrations 
like a fingerprint sensor that is not available on mobile platforms like iOS. 
In particular, the Session Controller within the USSD Layer manages session 
lifecycle and state transitions, coordinating with the Android Accessibility 
Services API to perform real-time background reading and automated interaction 
with USSD dialogs.

To meet the data privacy requirements of financial transactions, the system 
is designed as a fully on-device solution. This means that sensitive user data, 
such as voice inputs and transaction details, is processed locally without 
relying on external cloud servers. Voice commands are captured through the 
Speech Recognition module and interpreted by an Offline Natural Language 
Processing (NLP) Engine, ensuring functionality without dependence on 
cloud-based inference.

A key part of this deployment approach is the use of Android's 
hardware-backed Trusted Execution Environment (TEE). Instead of using custom 
software-based storage, which may be vulnerable, the system relies on the 
native Android Keystore. This ensures that the user's mobile money PIN is 
securely stored at the hardware level and can only be accessed after successful 
biometric authentication on the device.

The on-device system communicates outward with external infrastructure through 
the Mobile Network/USSD Gateway, which routes USSD requests to the Mobile 
Money System where transaction processing is handled. This outbound channel 
represents the only point of external dependency in the architecture.

By combining these native Android components including local Text-to-Speech 
(TTS), offline natural language processing, and the Keystore API the system 
provides a secure, private, and self-contained environment that works reliably 
even in low-resource settings.

\subsection{User Journey}
\begin{figure*}[h]
\centering
\includegraphics[width=\linewidth]{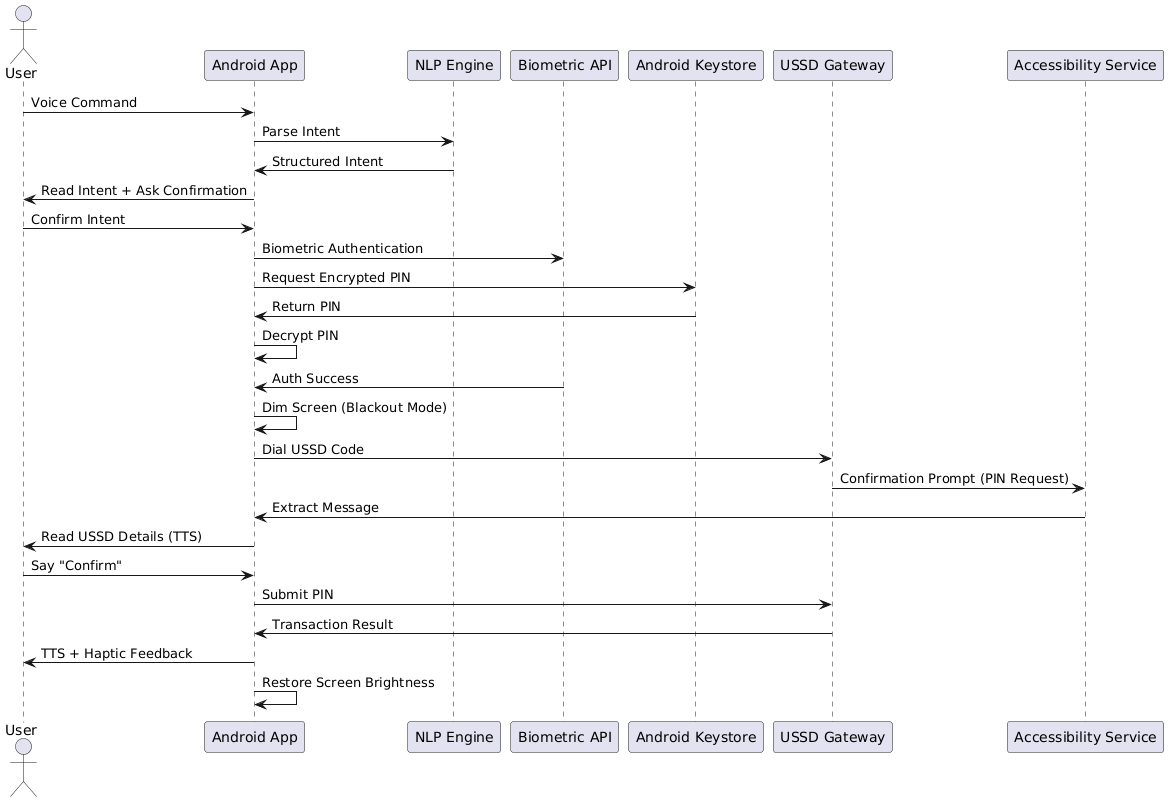}

\caption{Sequence Diagram for the transacation flow}
\label{fig:transaction_sequence}
\end{figure*}

This section describes the end-to-end interaction flow between the user and the proposed USSD automation system. The system is designed to support visually impaired users by enabling voice-driven interaction, secure authentication, and automated USSD navigation. 

Figure~\ref{fig:transaction_sequence} illustrates the high-level workflow of the system, beginning from voice input capture to transaction completion and feedback delivery. The process integrates on-device natural language processing, accessibility services, and secure credential handling to ensure both usability and security throughout the interaction.

\subsubsection{User Onboarding}

The system includes a structured onboarding process designed to configure user identity, network provider selection, and secure authentication setup before enabling full transaction functionality.

\begin{figure*}[h]
\centering
\includegraphics[width=\linewidth]{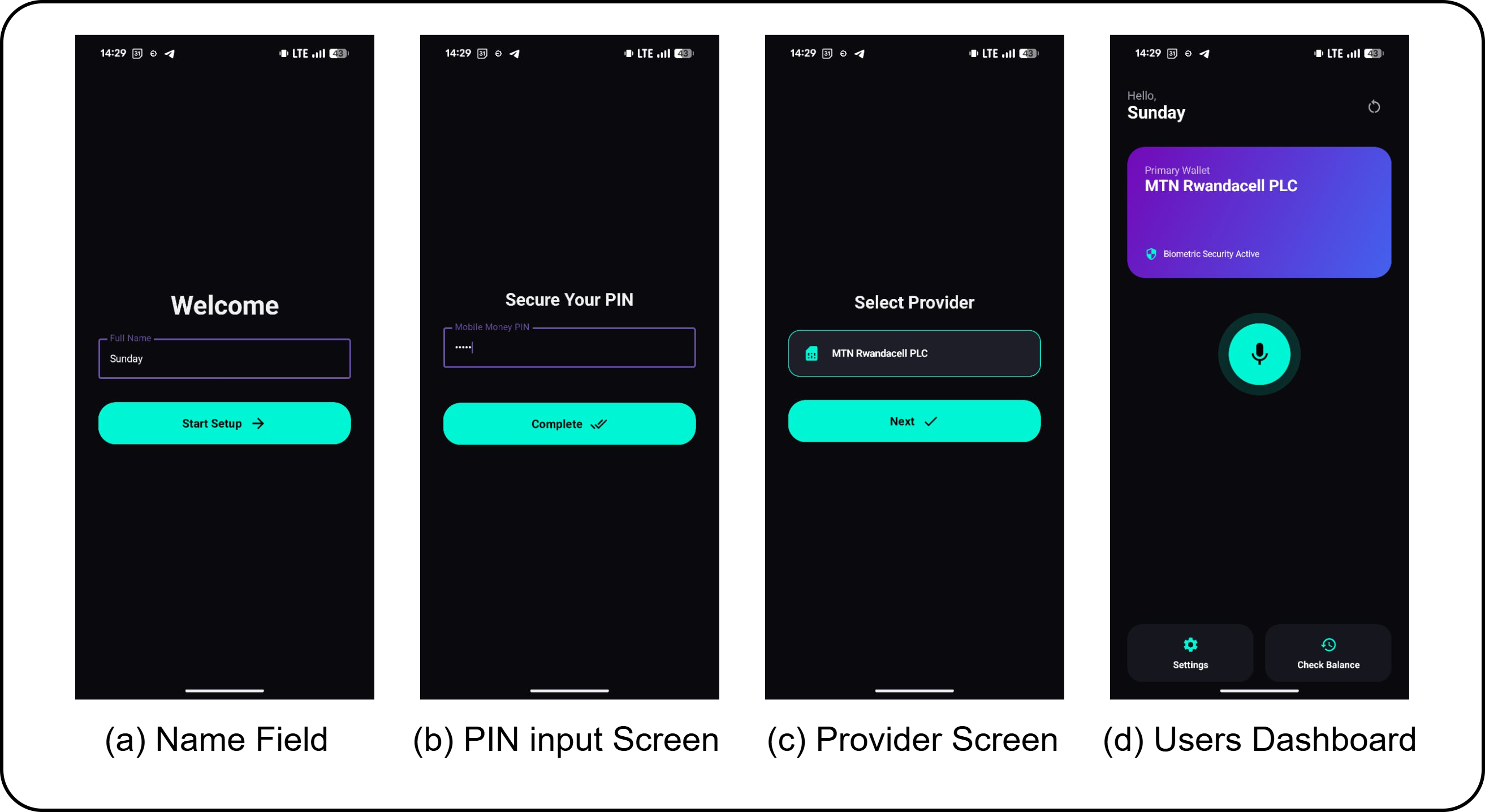}

\caption{User Onboarding flow}
\label{fig:user_flow}
\end{figure*}

\begin{itemize}

\item \emph{User Identity Registration.}
During initial setup, the system prompts the user to verbally state their name. The input is captured using the device microphone and processed using on-device speech recognition. The recognized name is stored locally and used to personalize subsequent interactions, including transaction confirmations and auditory feedback.

\item \emph{Telecom Service Provider Selection.}
In scenarios where multiple SIM cards are present on the device, the system detects available network operators using the TelephonyManager API. The user is then prompted to select their preferred mobile money provider (e.g., MTN, Safaricom, Airtel). This selection determines the USSD dialing prefix and transaction routing logic for all future operations.

\item \emph{Transaction PIN Setup.}
The final onboarding step involves secure PIN registration. The user is prompted to verbally confirm the creation of a transaction PIN, which is then entered using an accessibility-guided secure input interface. The PIN is immediately encrypted and stored within the Android Keystore (hardware-backed Trusted Execution Environment) to ensure that it is never stored in plaintext or exposed to external storage.

Biometric authentication is subsequently linked to the stored PIN, enabling secure authorization for all future transactions without requiring manual PIN re-entry.

\end{itemize}


\section{Validation}
To validate the proposed solution, we have run  experiments using three main scenarios, described in Section \ref{sec:usecases} Use cases, in USSD-based mobile money systems.
This validation phase was conducted by the development team, who simulated the interaction patterns of visually impaired users to assess the functionality of the system.

\subsection{Use cases}
\label{sec:usecases}
\begin{itemize}[]
\item \emph{Use Case 1: Check Wallet Balance}: This use case allows the user to check the balance in their mobile money account. Traditionally, to check the balance, the user dials a USSD code, for instance in Rwanda, to check balance with MTN, you use ``*182\#`` and navigates menu options using the phone interface or use ``*182*6*1\#``.
\item \emph{Use Case 2:Airtime Top-Up}: This use case allows users to recharge their mobile money accounts. For instance, using MTN Rwanda, you can dial ``*182\#`` and following the menu options or you use ``*182*2*1*1*1\#``

\item \emph{Use Case 3: Fund Transfer}: This use case allows the user to send money to a recipient by dialing the USSD code.  In Rwanda, using MTN, you can either send funds to a merchant code using ``*182*8*1*merchant code*amount\#`` or to a recipient number using ``*182*1*1*recipient number*amount\#``.
\end{itemize}

\subsection{Experiments}

The experiments focused  on the three cases and compared how mobile money transactions are  traditionally performed versus how they are executed using the proposed system. This comparison highlights accessibility challenges faced by visually impaired users and shows how the system improves usability, security, and efficiency. For each case, we maintained the high level implementation flow shown in figure \ref{fig:transaction_sequence}

\begin{table*}[h]
\centering
\caption{Comparison of Use Cases: Traditional vs Proposed System}
\label{tab:usecase_comparison}
\begin{tabularx}{\linewidth}{|l|X|X|}
\hline
\textbf{Use Case} & \textbf{Traditional (Screen Reader)} & \textbf{Proposed System} \\
\hline

Check Balance 
& In a traditional USSD balance-checking flow, the user uses a screen reader’s voice command to dial a code. The system presents multiple menu options, which are read aloud by the screen reader. The user navigates through these menus step by step, selecting the appropriate options until reaching the “check balance” function. Once selected, the system retrieves the balance and reads it back to the user via the screen reader.
& We use voice command triggers automated USSD session with TTS feedback, biometric authentication, and blackout mode which  enables a secure, fast, and fully non-visual interaction. \\
\hline

Airtime Top-Up 
& Using a screen reader application, the user issues a voice command to navigate through multiple USSD menu options to access the airtime top-up feature, with the system reading each step aloud. The user then manually enters the required amount and confirms the transaction, with all prompts and confirmations read back by the screen reader.
& User uses a voice command to specify amount, the system automates USSD flow with TTS feedback and secure PIN handling, enabling quick and independent execution. \\
\hline

Fund Transfer 
& Using a screen reader application, the user issues a voice command to dial the USSD code and navigates through multiple menu options to access the fund transfer feature. The system reads each menu option aloud, and the user responds verbally or manually selects the appropriate options, including entering the recipient’s phone number, specifying the transfer amount, and confirming the transaction. 
& User uses a voice command which  includes recipient and amount, system validates input, automates transaction, and ensures secure PIN entry with audio feedback and no visual interaction. \\
\hline

\end{tabularx}
\end{table*}

\section[Results and Discussion]{\uppercase{Results and Discussion}}
This section presents and discusses the results of our experiment.  
In all settings, the proposed system demonstrated the ability to:
\begin{inparaenum}[]
    \item Interpret voice commands into structured transaction intents;
    \item Automatically initiate USSD sessions through background dialing;
    \item Extract and parse USSD responses using Android Accessibility Services;
    \item Deliver real-time auditory feedback via Text-to-Speech (TTS);
    \item Trigger biometric authentication at the PIN entry stage;
    \item Securely inject encrypted PINs using the Android Keystore; and
    \item complete end-to-end transactions without requiring visual interaction.
\end{inparaenum}
This indicates that the system can successfully automate the full USSD transaction workflows while preserving both usability and security for visually impaired users.


\subsection{Results}
Results from experimentation with the three use cases, as shown in Table \ref{tab:operation_comparison}, indicate that the proposed system introduces significant improvements, particularly in terms of accessibility, security, efficiency, and reliability.

\begin{table*}[h]

\centering
\caption{Comparison of Traditional USSD and Proposed System}
\label{tab:operation_comparison}
\begin{tabularx}{\linewidth}{|l|X|X|}
\hline
\textbf{Feature} & \textbf{Traditional USSD (Screen Reader)} & \textbf{Proposed System} \\ 
\hline
Input Method & Manual dialing + screen reader navigation & Voice command interaction \\ 
\hline
Navigation & Sequential and time-consuming & Fully automated execution \\ 
\hline
Accessibility & Limited compatibility with screen readers & Fully non-visual (TTS + voice interaction) \\ 
\hline
PIN Entry & Manual entry, often requires assistance & Biometric authentication + secure PIN injection \\ 
\hline
Privacy & High risk due to third-party assistance & High (blackout mode + hardware-backed Keystore) \\ 
\hline
Transaction Time & 40–60 seconds & 12–15 seconds \\ 
\hline
Error Rate & High (manual input and navigation errors) & Low (automation reduces human error) \\ 
\hline
Independence & Low (dependency on assistance) & High (fully autonomous interaction) \\ 
\hline
\end{tabularx}
\end{table*}

\begin{itemize}
    \item \emph{Accessibility:} Eliminates the dependence on visual navigation and third-party assistance
    \item \emph{Security:} Protects sensitive information through biometric authentication and secure PIN handling
    \item \emph{Efficiency:} Reduces transaction time from approximately 40--60 seconds to 12--15 seconds
    \item \emph{Reliability:} Minimizes errors and session timeouts through automation
\end{itemize}

While the system is designed for visually impaired users, these improvements also benefit a broader user base by simplifying interaction, reducing cognitive load, and improving overall transaction speed.


\subsection{Discussion}

The system's performance was evaluated using controlled testing scenarios and compared against conventional manual USSD interaction.

\subsubsection{Time-to-Completion}

The proposed system significantly reduced transaction completion time:

\begin{itemize}
    \item Traditional USSD with screen reader: 40--60 seconds
    \item Proposed system: 12--15 seconds
\end{itemize}

This improvement is attributed to the following:\emph{
\begin{inparaenum}[]
    \item elimination of manual menu navigation;
    \item predefined transaction execution flows; and 
    \item faster system-driven confirmation handling
\end{inparaenum}}

\subsubsection{Task Success Rate}

The system achieved improved transaction reliability under normal network conditions: 

\begin{itemize}
    \item Traditional approach: 65--75\%
    \item Proposed system: more than 90\%
\end{itemize}

Observed failures were primarily caused by:
\begin{inparaenum}[]
    \item network latency and interruptions, and 
    \item variability in USSD response structures across providers
\end{inparaenum}

    \section[Limitations]{\uppercase{Limitations}}

\subsection{Voice Interaction in Noisy Environments}

The system relies on voice commands as the primary input for visually impaired users. However, speech recognition performance can be negatively affected in noisy real-world environments, such as markets, public transport, or crowded spaces.

Environmental noise, microphone quality, and speech clarity may affect the accuracy of command recognition, potentially leading to incorrect parsing of transaction details or the need for repeated input attempts. This limitation impacts usability in the very contexts where mobile money is most frequently used.

\subsection{Biometric Authentication Constraints}

The proposed solution uses biometric authentication (primarily fingerprint recognition) to securely authorize PIN release. While this enhances both usability and security, not all smartphones support biometric capabilities, particularly lower-end Android devices commonly used in low-resource settings.

In such cases, fallback authentication mechanisms (e.g., manual PIN entry) may be required, which could reintroduce accessibility challenges and partially undermine the system’s goal of eliminating reliance on visual input.

\subsection{Regulatory and Telecom Restrictions}

The proposed system operates as a middleware layer over existing USSD infrastructure provided by telecom operators such as MTN, Safaricom, and Airtel. However, these operators do not officially expose programmable USSD APIs for third-party applications.

As a result, the system depends on indirect interaction mechanisms, which introduces potential compatibility risks with future network or operating system updates. Changes in USSD menu structures, response formats, or session handling by telecom providers may disrupt the automation workflow and require continuous adaptation. This lack of formal integration limits long-term stability and scalability.

\subsection{NLP Accuracy Constraints}

The system currently employs on-device Natural Language Processing (NLP) for intent recognition, with support limited primarily to English-language commands. This creates accessibility barriers for users who are more comfortable with local languages.

Additionally, NLP performance may be affected by:
\begin{itemize}
    \item Variations in accents and pronunciation
    \item Informal or ambiguous phrasing
    \item Differences in Speech Patterns Between Users
\end{itemize}
These factors can lead to misinterpretation of transaction intent, requiring additional confirmation steps and potentially affecting overall efficiency.
\section[Conclusion and Future work]{\uppercase{Conclusion and Future Work}}

This research introduced a secure and automated USSD middleware designed to enable visually impaired users to independently perform mobile money transactions without external assistance.

The system addresses a critical gap in digital financial accessibility by eliminating the dependency on third-party support, which often exposes users to privacy and fraud risks.
By integrating Android Accessibility Services, biometric authentication, hardware-backed encryption through the Android Keystore, and privacy-preserving screen dimming, the proposed solution effectively resolves the long standing security accessibility tradeoff in USSD-based financial systems.

The experimental evaluation demonstrated improved performance in terms of task success rate, transaction time, and user privacy compared to conventional manual USSD interaction. In particular, the proposed system increased task success rates from 65–75\% to over 90\% and reduced transaction completion time from 40–60 seconds to 12–15 seconds, while also improving perceived security among users.

In general, this work contributes to the intersection of FinTech security, accessibility engineering, and inclusive digital infrastructure, offering a practical and scalable approach to improve financial inclusion among visually impaired users in low-resource environments.

\subsection[Future Work]{\uppercase{Future Work}}

Future work will focus on improving accessibility, functionality, and system intelligence. This includes integrating multilingual NLP support (e.g., Kinyarwanda) to enable more inclusive voice interaction and expanding the platform to support additional financial services such as bill payments, savings, and transaction history.

Further improvements will incorporate adaptive and learning-based capabilities to better handle diverse USSD flows, personalize interactions, and enable predictive automation. In addition, adaptive session management techniques will be explored to mitigate USSD time constraints and reduce transaction failures.


\balance

\bibliographystyle{apalike}
{\small
\bibliography{references}} 

\end{document}